\documentclass[a4paper]{spie}

\usepackage{amsmath,amsfonts,amssymb}
\usepackage[]{graphicx}
\usepackage[]{color}
\usepackage{alltt}
\usepackage[T1]{fontenc}
\usepackage[utf8]{inputenc}
\usepackage[colorlinks=true, allcolors=blue]{hyperref}
\usepackage{cleveref}
\usepackage{siunitx}
\usepackage{xcolor}
\usepackage{blkarray}
\usepackage{subcaption}
\usepackage{capt-of}
\usepackage{listings}
\usepackage{enumitem}
\usepackage{comment}
\usepackage{algorithm}
\usepackage{algpseudocode}
\usepackage{makecell}
\usepackage[super]{nth}
\usepackage{tikz}
\usetikzlibrary{shapes,arrows,calc,fit}

\tikzstyle{block} = [draw, rectangle, minimum height=3em, minimum width=6em]
\tikzstyle{sum} = [draw, circle, node distance=1cm]
\tikzstyle{input} = [coordinate]
\tikzstyle{output} = [coordinate]
\tikzstyle{pinstyle} = [pin edge={to-,thin,black}]

\algnewcommand\algorithmicto{\textbf{to}}
\algnewcommand\RETURN{\State \textbf{return} }

\providecommand{\keywords}[1]{\textbf{\textit{Keywords:}} #1}

\title{DD4AO control law for RISTRETTO: robustness, real-time performance, and on-sky validation with PAPYRUS}

\author[a]{Isaac Dinis}
\author[a]{Muskan Schinde}
\author[a]{Nicolas Blind}
\author[a]{Bruno Chazelas}
\author[a]{Nathanael Restori}
\author[a]{Christophe Lovis}
\author[b]{Sylvain Cetre}
\author[c]{Angelie Alagao}
\author[c]{Romain Fetick}
\author[c]{Taïssir Héritier}
\author[c]{Vincent Chambouleyron}
\author[c]{Benoit Neichel}
\author[d]{Thierry Fusco}
\author[d]{Jean-François Sauvage}

\affil[a]{Univérsité de Genève (UNIGE), Département Astronomique, Geneva, Switzerland}
\affil[b]{Wakea consulting, France}
\affil[c]{Laboratoire d'Astrophysique de Marseille (LAM), Marseille, France}
\affil[d]{Office national d'études et de recherches aérospatiales (ONERA), France}

\authorinfo{Further author information: Send correspondence to Isaac Dinis : isaac.dinis@unige.ch}

\begin{document} 
\maketitle

\begin{abstract}
This study presents DD4AO progress towards its implementation in the RISTRETTO\cite{ristretto} 
instrument. DD4AO is a novel frequency-domain, data-driven controller for adaptive 
optics that leverages power spectral density estimation for optimization while 
enforcing stability criteria. It addresses disturbance rejection, command amplitude 
constraints, and system transfer functions through convex optimization, yielding an 
optimal controller in Infinite Impulse Response (IIR) filter form. We present the 
on-sky validation of DD4AO conducted using the PAPYRUS\cite{papyrus} instrument at the 
Observatoire de Haute-Provence (OHP). The observations were performed on two stars 
over the night of 24--25 March 2026: the bright star Arcturus, and the faint binary 
HD137909. DD4AO successfully maintained a closed and stable loop over hour-long 
exposures while continuously adapting to evolving atmospheric conditions. The 
pipeline enabled instantaneous switching between DD4AO and standard controllers,
namely the Integrator and OMGI, allowing direct statistical comparisons 
throughout each observation. On Arcturus, DD4AO achieved a 5\% Strehl ratio 
improvement over the integrator at $\lambda = 1310\,\text{nm}$, from 28.8\% to 
33.9\%. On HD137909, performance differences were smaller due to the low-SNR 
regime, though DD4AO consistently used less deformable mirror stroke and 
suppressed vibration peaks present in the residuals of the standard controllers. 
These results validate DD4AO as a robust on-sky control solution and represent an 
important milestone towards its deployment in RISTRETTO and SAXO+\cite{SAXO+} at the VLT.
\end{abstract}


\section{INTRODUCTION}
\label{sec:intro}

Adaptive Optics (AO) is a key technology for ground-based astronomical imaging, enabling real-time correction of wavefront distortions introduced by atmospheric turbulence. Since its inception, AO has undergone continuous development, with eXtreme Adaptive Optics (XAO) representing its most advanced implementation. A prime example is the SPHERE instrument \cite{SPHERE} at the Very Large Telescope (VLT), which delivers high-contrast imaging through a high-order, high-speed AO system.

The next generation of XAO instruments at the VLT includes RISTRETTO and SAXO+. Both instruments aim to push AO performance further, both through hardware upgrades such as, more sensitive wavefront sensors (pyramid WFS) and faster deformable mirrors, and through advances in control software, particularly the adoption of advanced control laws. DD4AO is a novel frequency-domain, data-driven controller being developed for deployment in both of these next-generation XAO instruments.

As part of RISTRETTO's development, a collaboration with the PAPYRUS team at the Observatoire de Haute-Provence has enabled on-sky testing of several key RISTRETTO subsystems, including the unmodulated pyramid wavefront sensor, the spectrograph, and the DD4AO control law. This paper reports on the on-sky validation of DD4AO within this framework, demonstrating its performance and stability under real observing conditions.

\section{DD4AO: PRINCIPLE}

DD4AO is based on the work of Karimi et al.\ \cite{Karimi_2017}. It synthesizes 
an Infinite Impulse Response (IIR) filter of a specified order directly from 
non-parametric frequency response magnitudes, using mixed-sensitivity criteria 
combining $\mathcal{H}_2$ and $\mathcal{H}_\infty$ norms. For an AO 
implementation, the relevant frequency responses are the system transfer function, typically simplified as a pure delay, and the disturbance frequency 
magnitude, obtained from the periodogram of the pseudo open-loop reconstruction.

The controller design is formulated as a convex optimization problem and solved 
using CVXPY \cite{diamond2016cvxpy, agrawal2018rewriting}, with Clarabel 
\cite{Clarabel_2024} as the solver for the resulting Second-Order Cone Programming 
(SOCP) problem. Stability constraints, disturbance rejection, measurement noise 
mitigation, and actuator stroke penalties are all incorporated into the 
optimization. The optimization objective can be summarized as:

\begin{equation}
    \min_{K(z)}\Big( \left\|\Phi(z)\,\mathcal{S}(z)\right\|_2^2 + 
    \sigma(z)\left\|\mathcal{T}(z)\right\|_\infty^2\Big)
\end{equation}

where $K(z)$ is the controller, $\Phi(z)$ is the disturbance frequency magnitude, 
$\mathcal{S}(z)$ is the sensitivity function of the closed-loop system, 
$\mathcal{T}(z)$ is the complementary sensitivity function, and $\sigma(z)$ is a 
sigmoid weighting function that penalizes large stroke commands at high frequencies.

\begin{figure}[H]
\begin{minipage}[b]{0.5\linewidth}
\centering
\includegraphics[width=7cm]{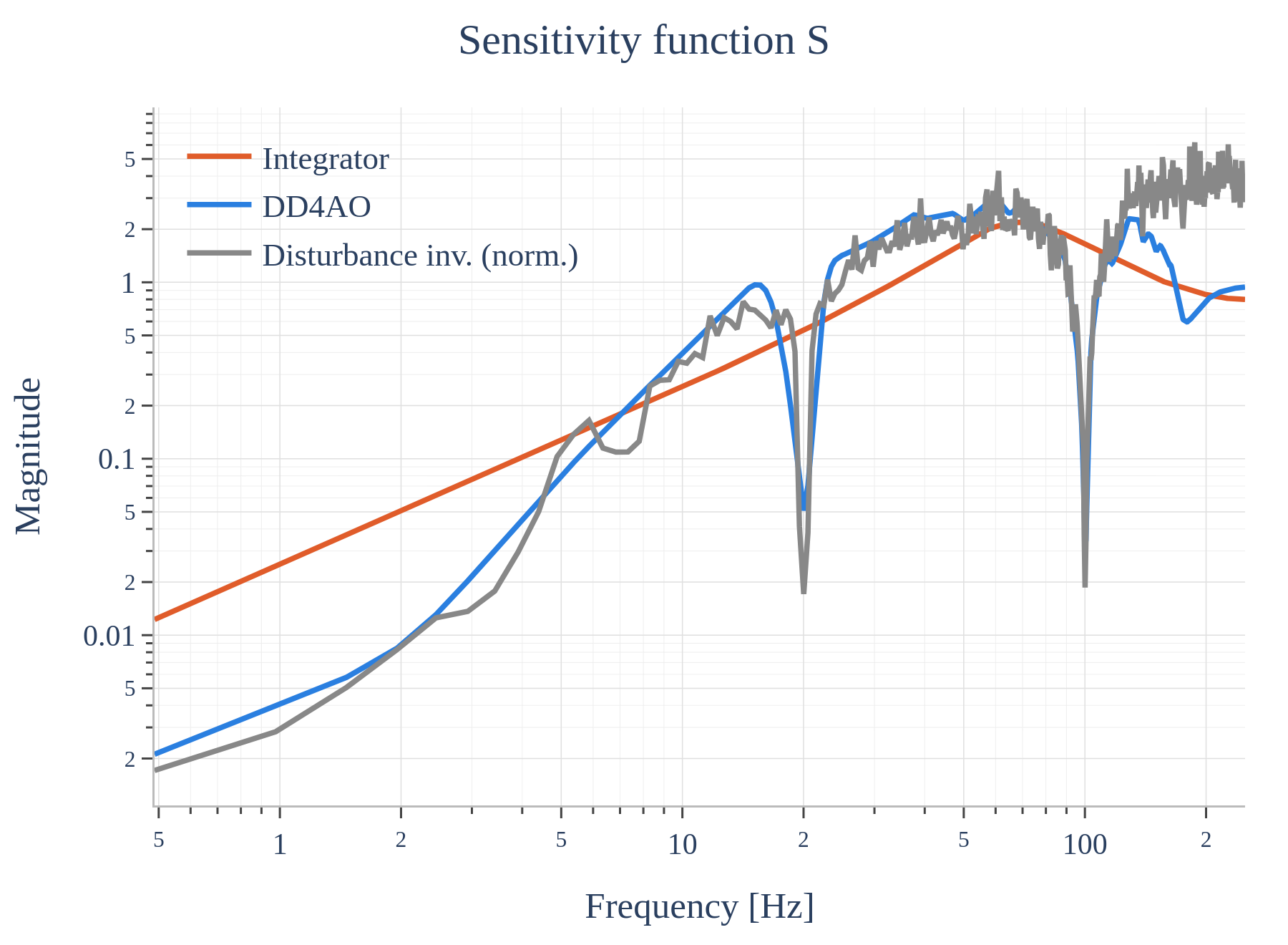}
\captionsetup{width=8cm}
\caption{DD4AO and integrator sensitivity function comparison with vibration peaks.}
\label{fig:sens_psd}
\end{minipage}
\begin{minipage}[b]{0.5\linewidth}
\centering
\includegraphics[width=7cm]{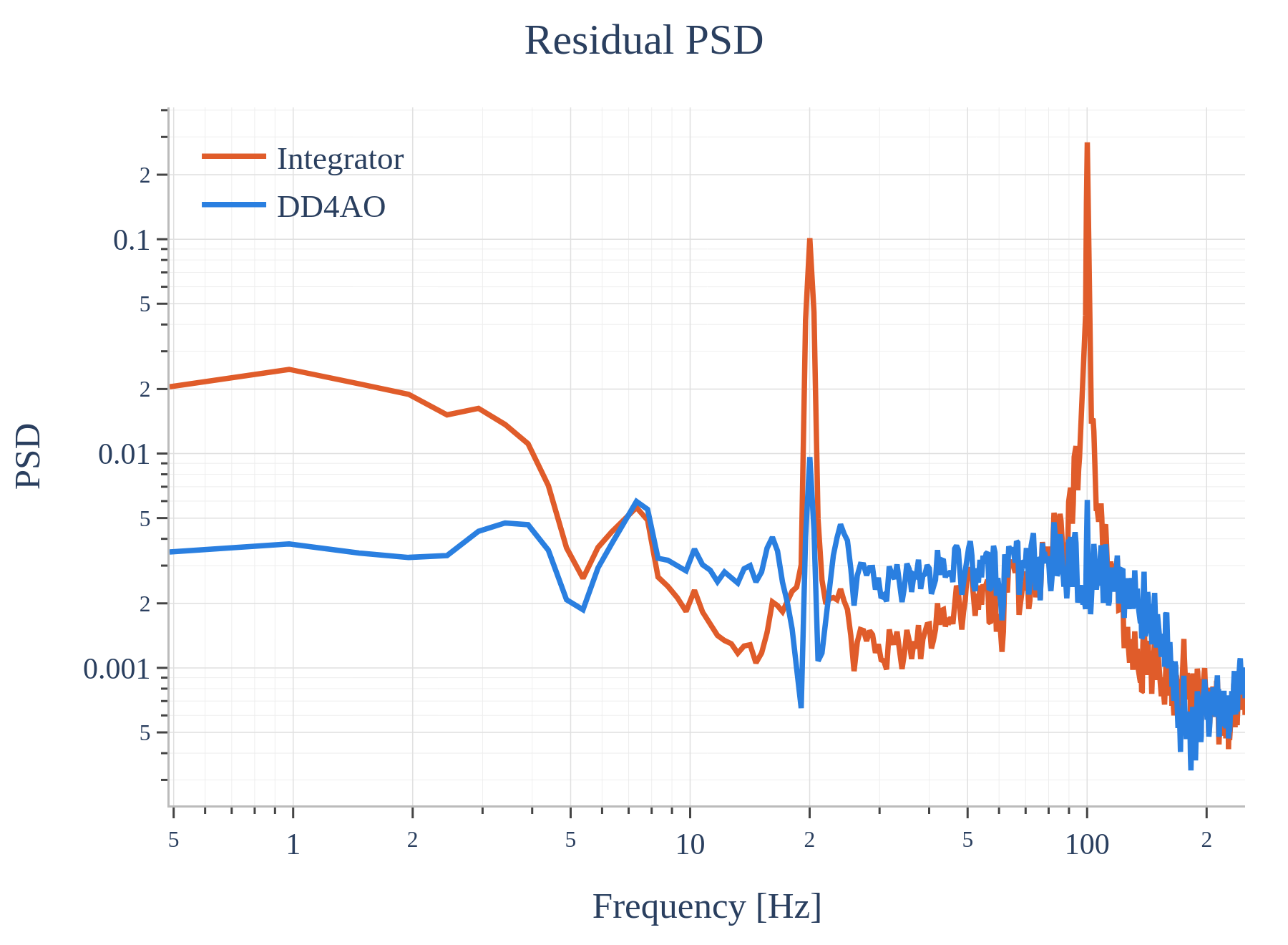}
\captionsetup{width=8cm}
\caption{DD4AO and integrator residual spectrum comparison with vibration peaks.}
\label{fig:rms_psd}
\end{minipage}
\end{figure}

\begin{figure}[H]
\begin{minipage}[b]{0.5\linewidth}
\centering
\includegraphics[width=7cm]{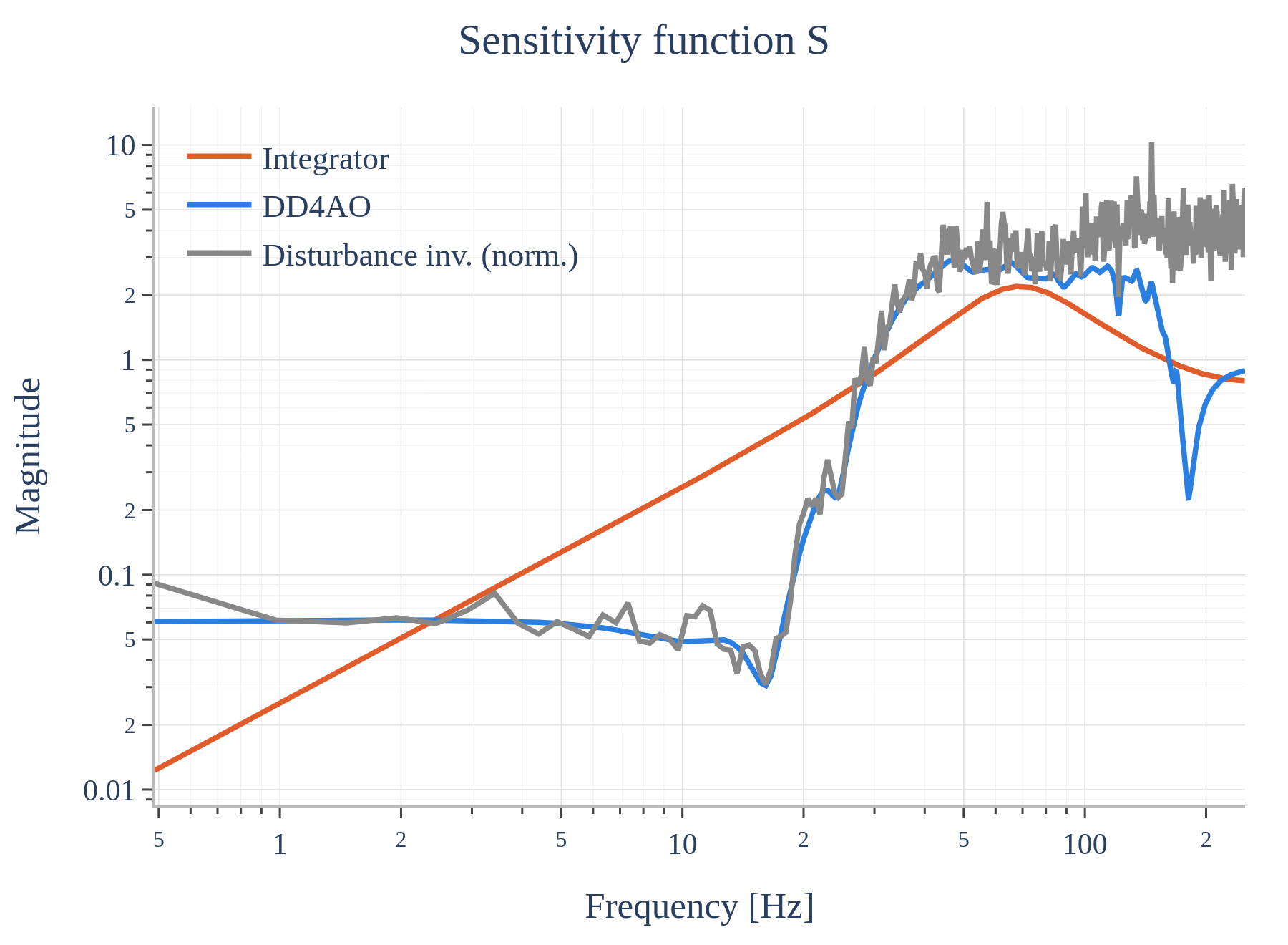}
\captionsetup{width=8cm}
\caption{DD4AO and integrator sensitivity function comparison with fast disturbance and low SNR.}
\label{fig:sens_psd2}
\end{minipage}
\begin{minipage}[b]{0.5\linewidth}
\centering
\includegraphics[width=7cm]{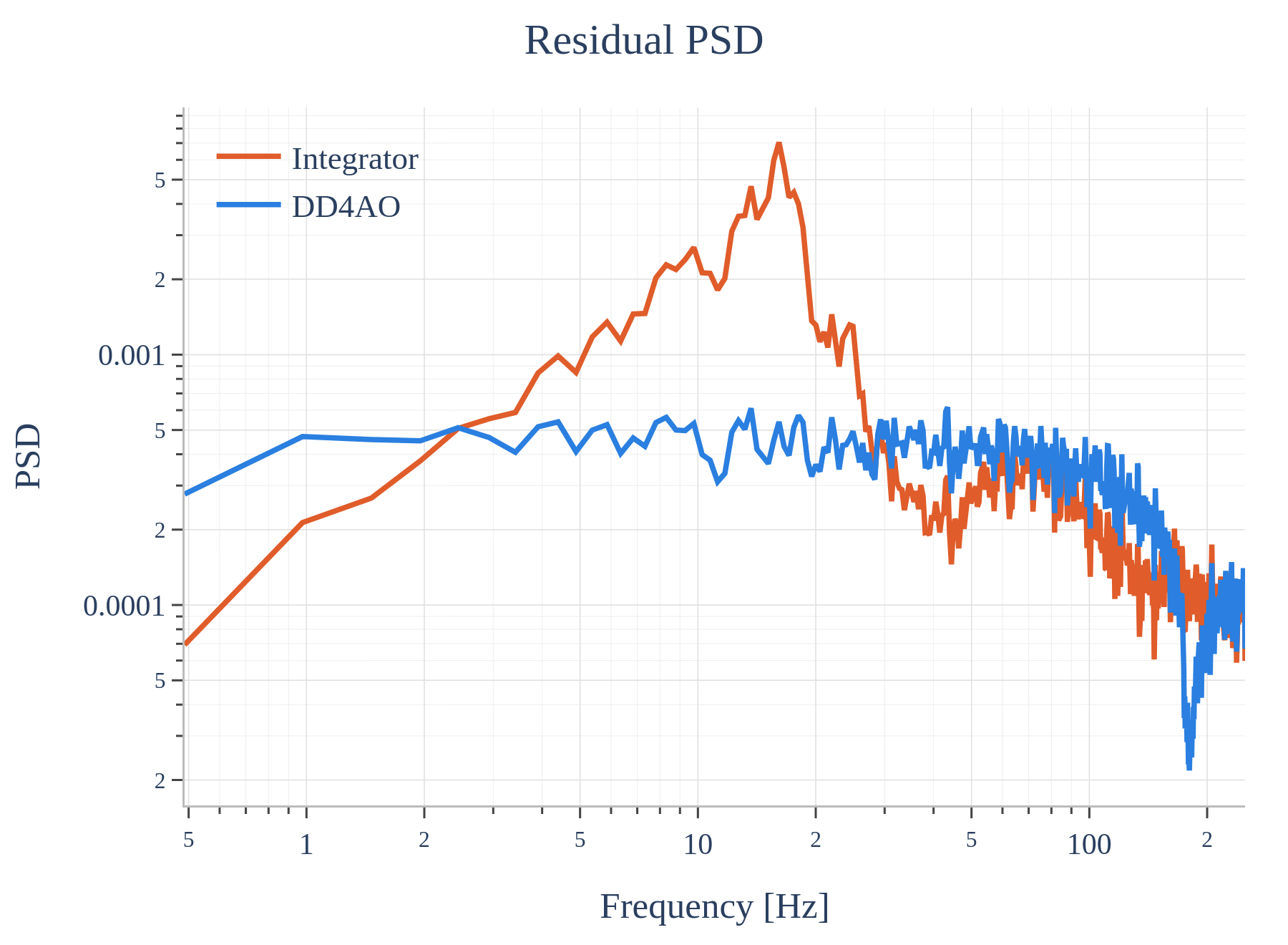}
\captionsetup{width=8cm}
\caption{DD4AO and integrator residual spectrum comparison with fast disturbance and low SNR.}
\label{fig:rms_psd2}
\end{minipage}
\end{figure}

Figures~\ref{fig:sens_psd},~\ref{fig:rms_psd},~\ref{fig:sens_psd2} and~\ref{fig:rms_psd2} illustrate the DD4AO optimized 
sensitivity function compared to that of a classical integrator, overlaid on the 
inverse of the disturbance frequency magnitude (left), along with the corresponding 
residual spectra (right). The DD4AO sensitivity (or rejection) function closely follows the 
disturbance profile, including vibration peaks, resulting in a notably flat 
residual spectrum, in contrast to the integrator.

\section{DD4AO REAL-TIME PIPELINE}

The DD4AO real-time pipeline has been implemented using the DAO RTC \cite{DAO}, 
using modal Karhunen-Lo\`eve control. It consists of three main processes:

\begin{enumerate}
    \item \textbf{Hard Real-Time Control:} This process applies the IIR filters 
    to the current measurements and past commands to compute the new command. As 
    this process is latency-sensitive, it is implemented in C. The measured latency 
    for 195 controlled modes and a filter order of 20 was \SI{0.04}{ms}. A similar 
    latency was measured using a simple integrator, as the dominant contribution 
    comes from memory transfers rather than the matrix-vector multiplication (MVM).

    \item \textbf{Open-Loop Reconstructor:} This process combines measurements and 
    commands to reconstruct the input disturbance of the AO system. It is a 
    soft real-time process running in Python.

    \item \textbf{Optimizer:} This process runs the convex optimization to compute 
    the IIR filters. It is triggered every 20 seconds to continuously adapt to 
    changing atmospheric conditions and vibrations, and is also implemented as a 
    soft real-time process in Python. Currently, optimizing a filter for a single 
    mode takes approximately 0.5 seconds. To keep the pipeline scalable to a large 
    number of controlled modes, filters are computed for groups of modes with 
    similar frequency magnitude profiles by averaging their periodograms. This 
    grouping strategy also improved robustness against noise and periodogram 
    estimation artifacts. Discussions with the CVXPY development team suggest that 
    the optimization runtime could potentially be reduced to a few milliseconds, 
    which will be investigated in future work, though it is not critical for the 
    current implementation.
\end{enumerate}

\begin{figure}[H]
\begin{center}
\begin{tabular}{c}
\includegraphics[height=5cm]{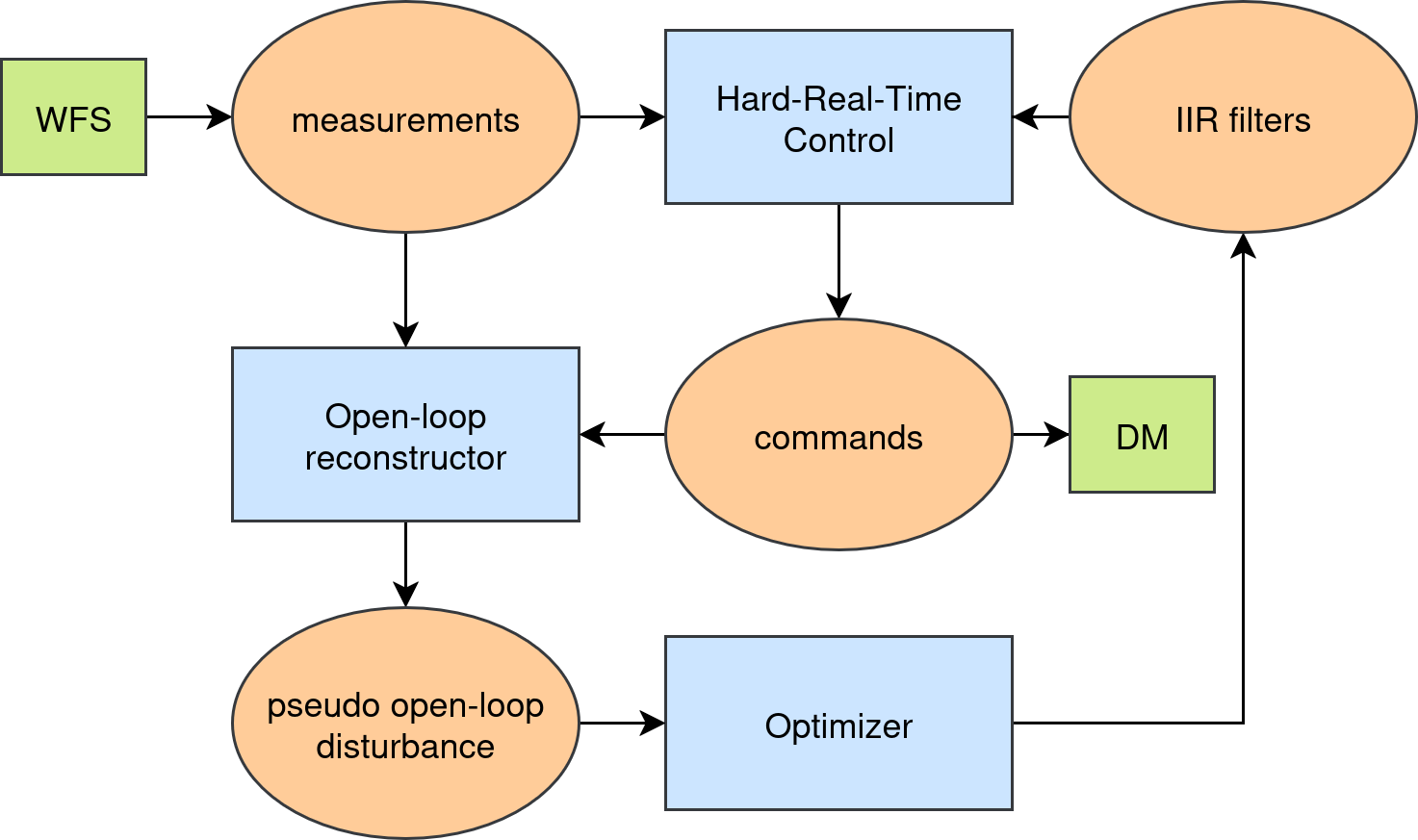}
\end{tabular}
\end{center}
\caption{Real-time pipeline of DD4AO.}
\label{fig:pipeline}
\end{figure}

Figure~\ref{fig:pipeline} shows a schematic of the full real-time pipeline.

\section{ON-SKY COMPARISON METHOD}

On-sky comparison of control laws is inherently challenging, as the uncontrolled 
nature of the environment makes it impossible to expose each controller to identical 
input turbulence. To address this, the real-time pipeline was designed to switch 
between control laws instantaneously while keeping the loop closed, so as not to 
disturb the simultaneous spectrograph testing. This allows switching back and forth 
between controllers and performing a statistical comparison over the full exposure.

During each observation, 15-second snapshots are taken for each control law within 
a 1-minute cycle, ensuring that all controllers sample similar atmospheric 
conditions over time.

One complication encountered was a significant optical gain. To 
compensate, a scalar gain was applied to all modal residuals and hand-tuned, with 
the same value used across all three control laws. While this made all three 
controllers operate sub-optimally, the performance penalty was applied equally, 
preserving the fairness of the comparison.

The observations were performed on two stars over the night of 24-25 March 2026: 
the bright star Arcturus (mag. 0), and the faint binary HD137909 (mag. 3.7).

\section{RESULTS}

\subsection{Arcturus}

\begin{figure}[H]
\begin{minipage}[b]{0.5\linewidth}
\centering
\includegraphics[width=8cm]{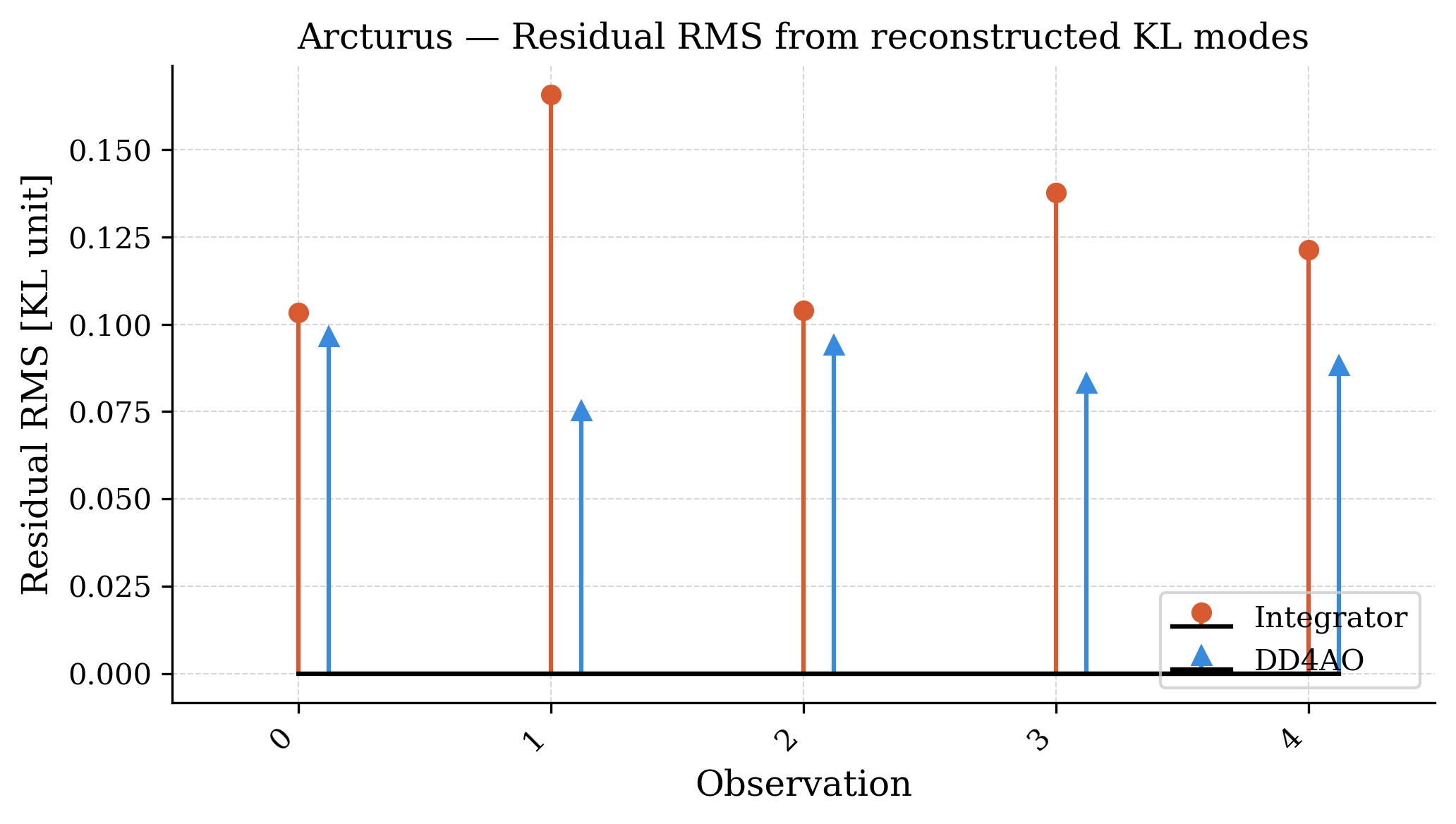}
\captionsetup{width=8cm}
\caption{Residual RMS from reconstructed KL modes across 5 Arcturus observations, 
comparing DD4AO and the integrator.}
\label{fig:arc_res}
\end{minipage}
\begin{minipage}[b]{0.5\linewidth}
\centering
\includegraphics[width=8cm]{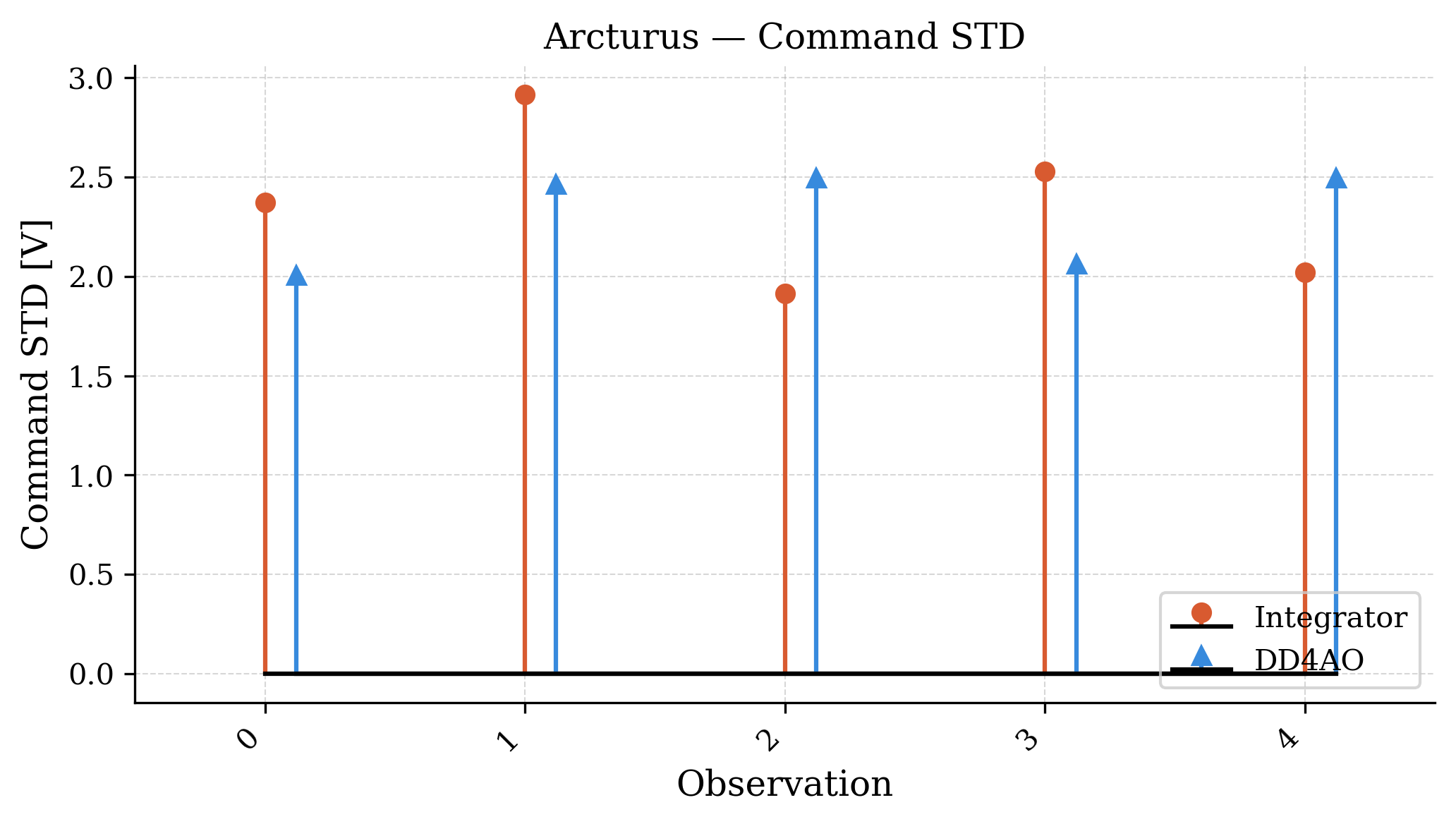}
\captionsetup{width=8cm}
\caption{Total residual PSD for a representative Arcturus exposure, comparing 
DD4AO and the integrator.}
\label{fig:arc_cmd}
\end{minipage}
\end{figure}

Figure~\ref{fig:arc_res} shows the residual RMS of the reconstructed KL modes for five observations taken within a one-hour period. DD4AO consistently outperforms the integrator, although the 
margin is not always statistically significant. This is likely attributable to the 
adaptive nature of DD4AO, which continuously adjusts to the evolving disturbance, 
whereas the integrator gain remained fixed and was not always optimal.
Figure~\ref{fig:arc_cmd} shows the command standard deviation in volts for the same 
observations. The two algorithms yield statistically similar command levels, 
confirming that they experienced comparable atmospheric disturbance throughout.

\begin{figure}[H]
\begin{minipage}[b]{0.5\linewidth}
\centering
\includegraphics[width=8cm]{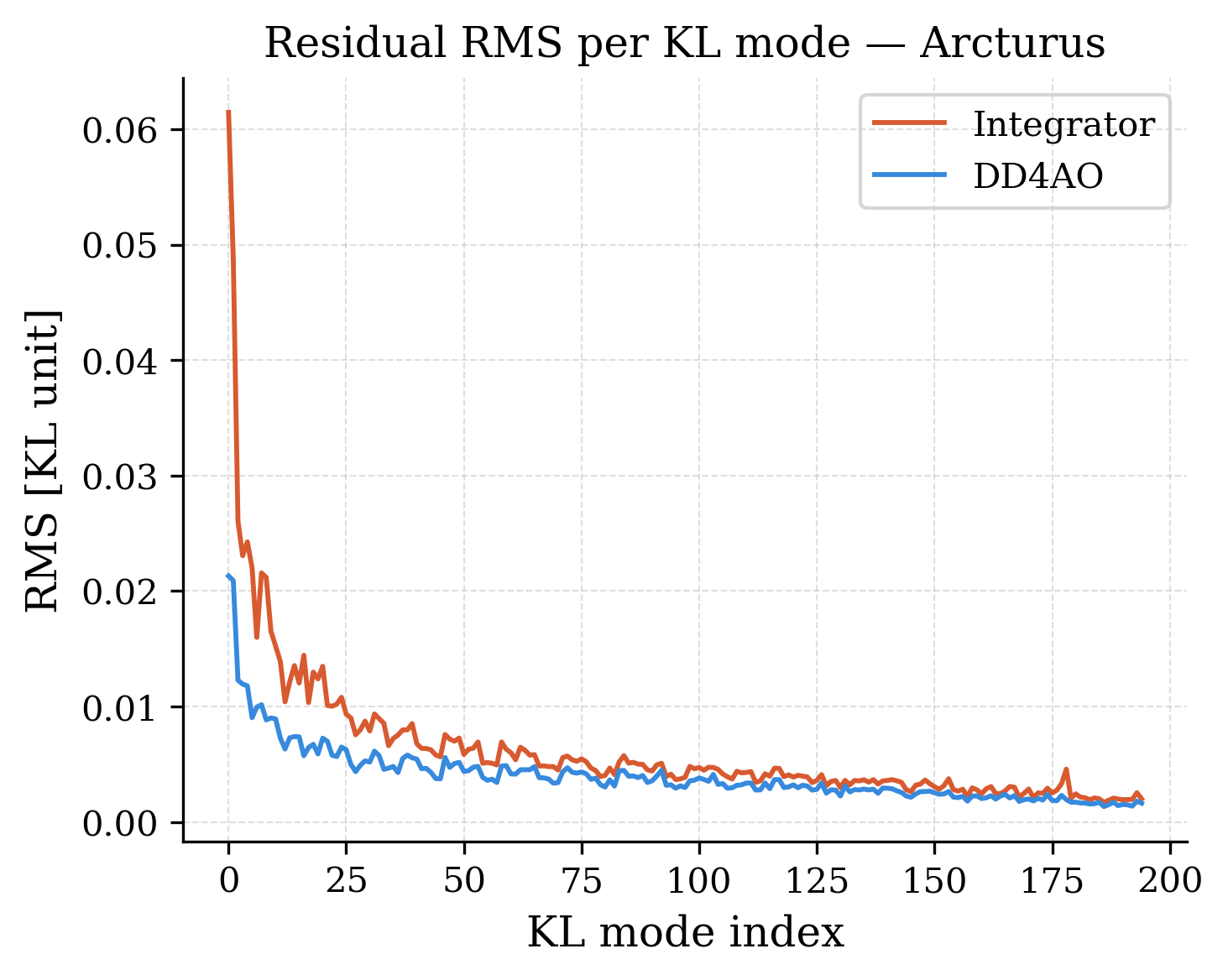}
\captionsetup{width=8cm}
\caption{Residual RMS per KL mode for a representative Arcturus exposure, 
comparing DD4AO and the integrator.}
\label{fig:arc_rms_mode}
\end{minipage}
\begin{minipage}[b]{0.5\linewidth}
\centering
\includegraphics[width=8cm]{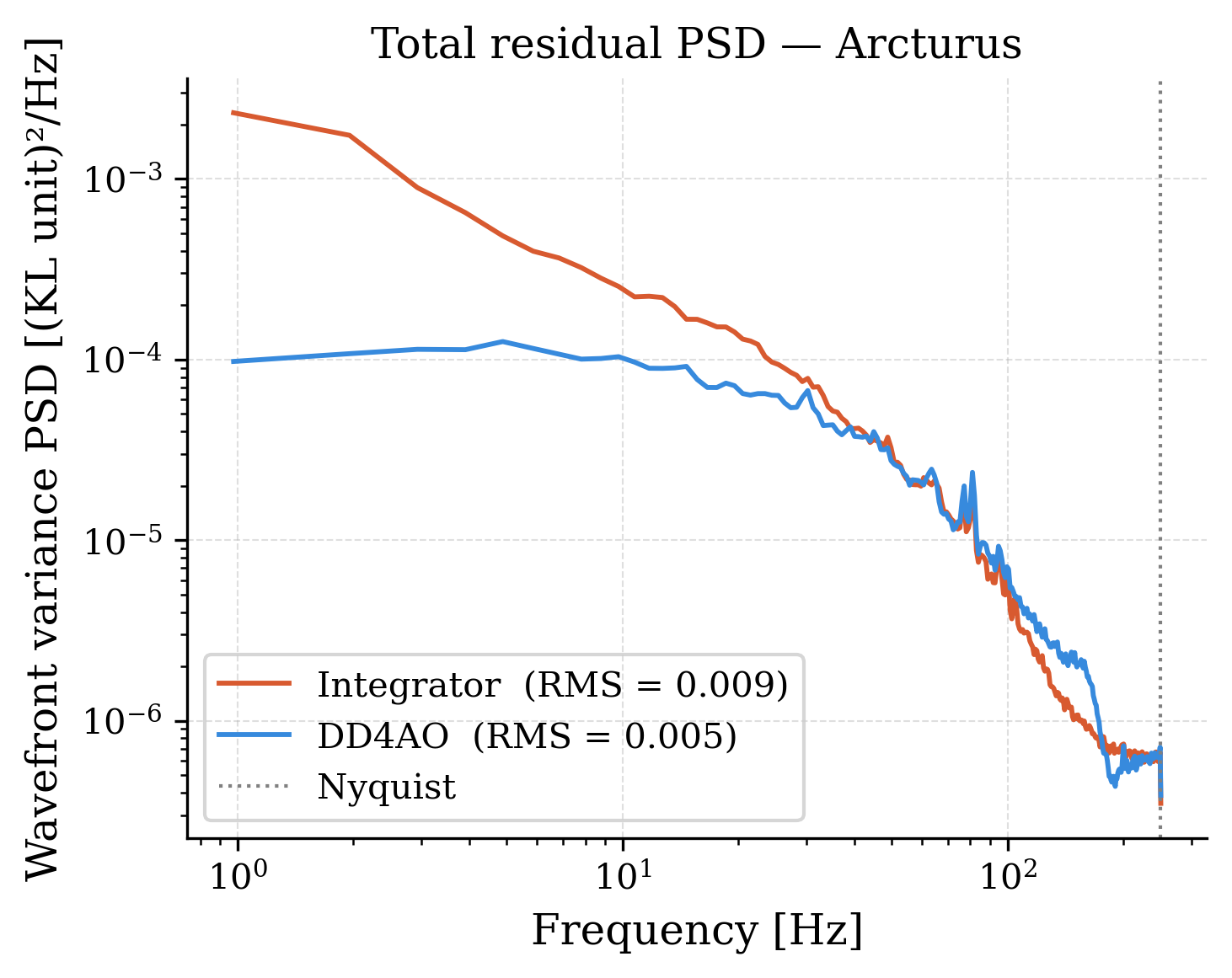}
\captionsetup{width=8cm}
\caption{Total residual PSD for a representative Arcturus exposure, comparing 
DD4AO and the integrator.}
\label{fig:arc_psd}
\end{minipage}
\end{figure}

Figures~\ref{fig:arc_rms_mode} and~\ref{fig:arc_psd} show, for one representative 
observation, the residual RMS per mode and the total residual PSD, respectively. 
The performance improvement brought by DD4AO is concentrated in the low-order 
modes, which is corroborated by the PSD plot, where the difference between the two 
controllers is confined to low temporal frequencies, consistent with low-order 
modal content. The PSD plot also suggests that the integrator gain may have been 
set too low, given the elevated low-frequency residuals and the absence of the 
typical integrator overshoot.

\begin{figure}[H]
\begin{minipage}[b]{0.5\linewidth}
\centering
\includegraphics[width=8cm]{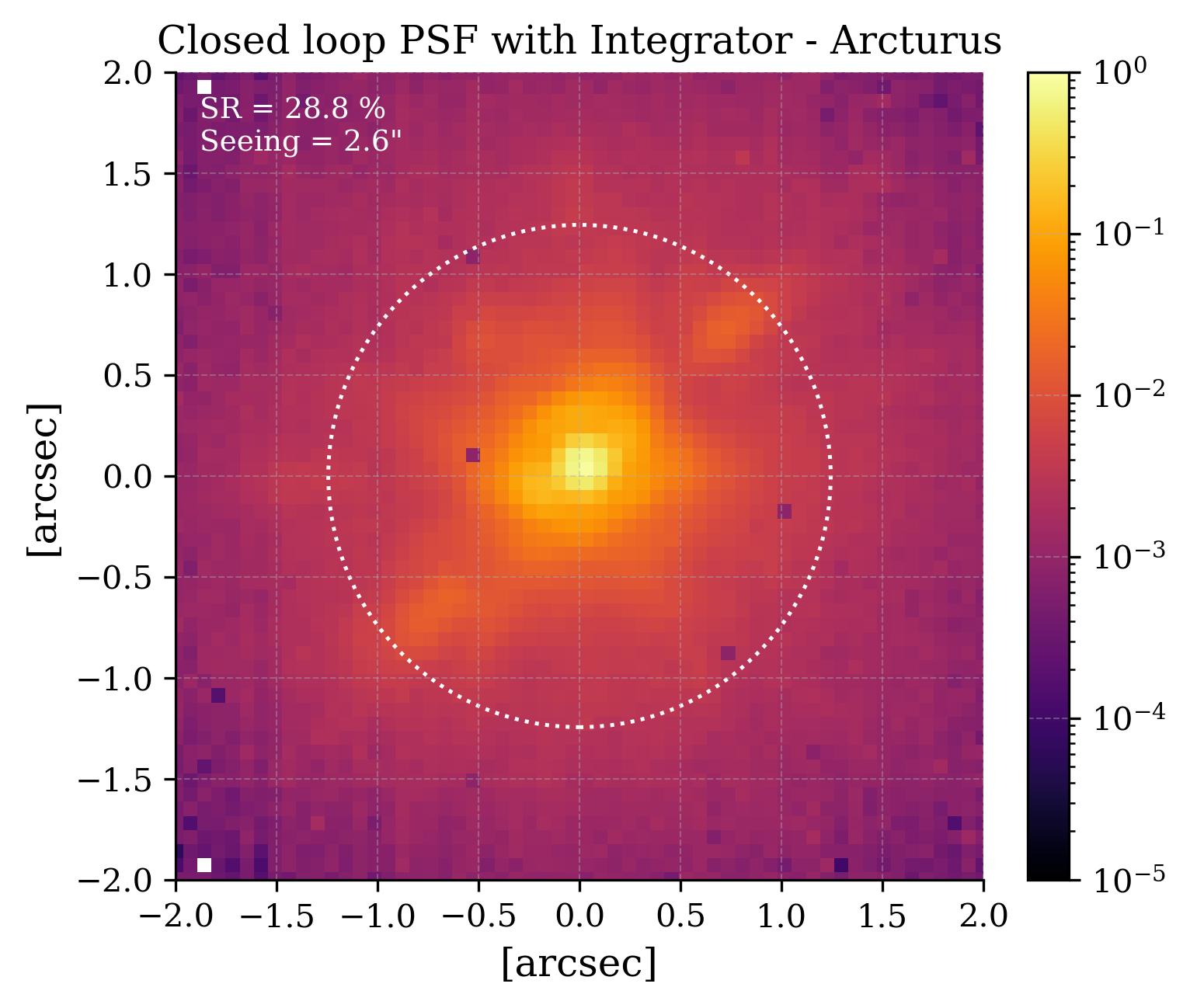}
\captionsetup{width=8cm}
\caption{Closed-loop PSF on Arcturus using the integrator. Strehl ratio at 
$\lambda = 1310\,\text{nm}$: 28.8\%.}
\label{fig:psf_int_arc}
\end{minipage}
\begin{minipage}[b]{0.5\linewidth}
\centering
\includegraphics[width=8cm]{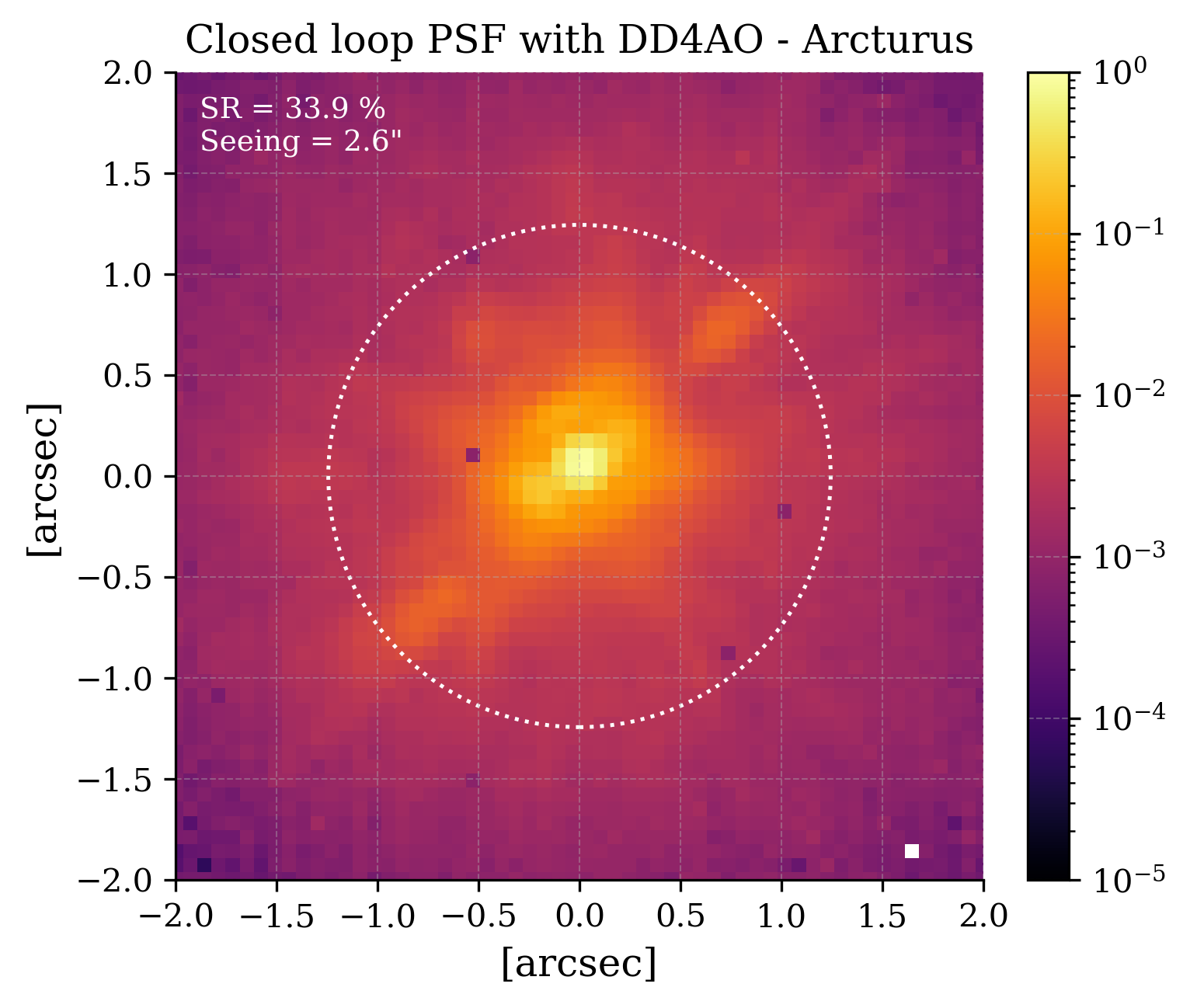}
\captionsetup{width=8cm}
\caption{Closed-loop PSF on Arcturus using DD4AO. Strehl ratio at 
$\lambda = 1310\,\text{nm}$: 33.9\%.}
\label{fig:psf_dd4ao_arc}
\end{minipage}
\end{figure}

Finally, Figures~\ref{fig:psf_int_arc} and~\ref{fig:psf_dd4ao_arc} show the 
closed-loop PSFs at $\lambda = 1310\,\text{nm}$ for a representative observation, 
demonstrating a 5\% Strehl ratio improvement with DD4AO. The strehl ratio is estimated with maoppy \cite{maoppy}.

\subsection{HD137909}

\begin{figure}[H]
\begin{minipage}[b]{0.5\linewidth}
\centering
\includegraphics[width=8cm]{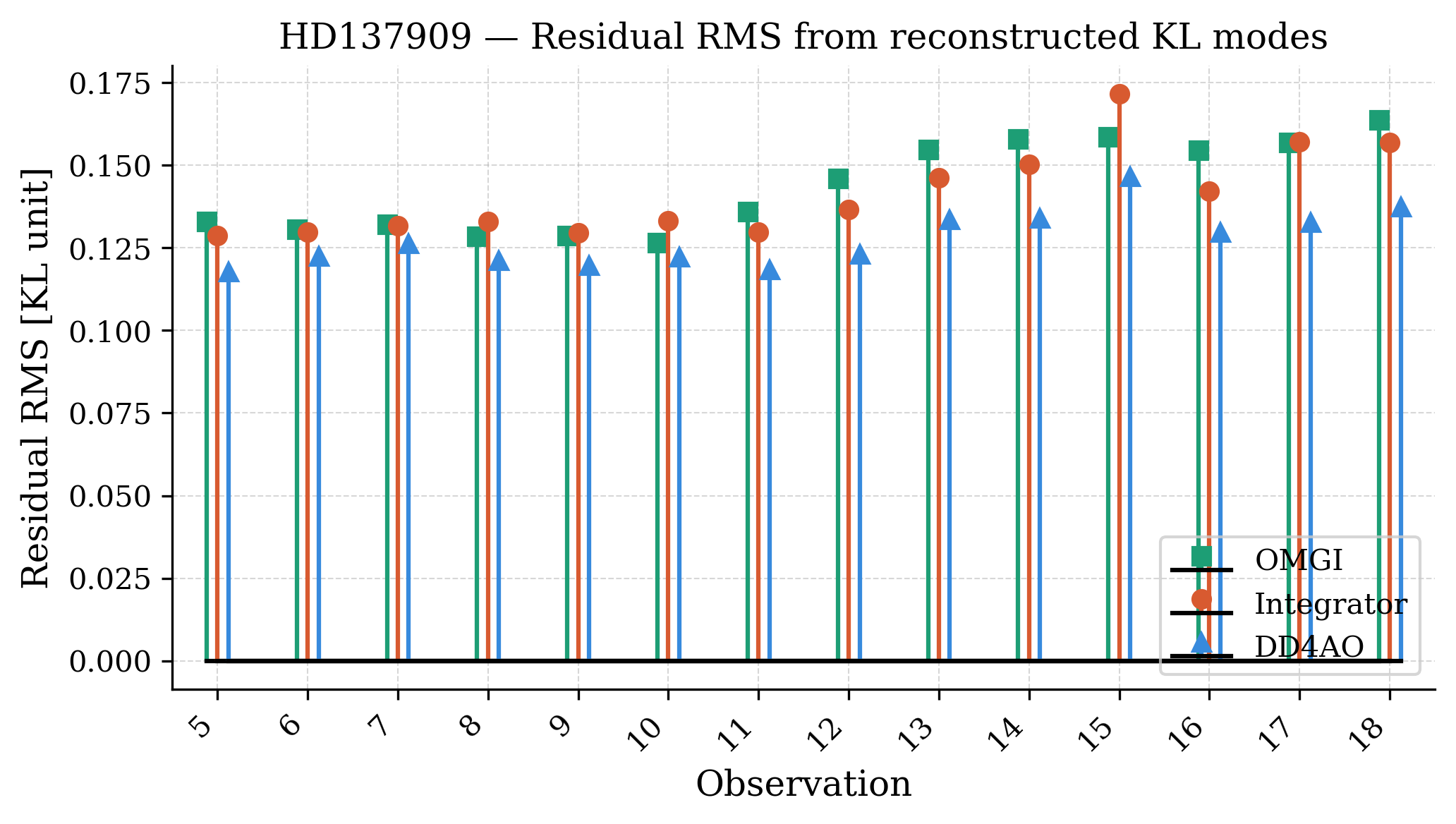}
\captionsetup{width=8cm}
\caption{Residual RMS from reconstructed KL modes across 14 HD137909 observations, 
comparing DD4AO and the integrator.}
\label{fig:hd_res}
\end{minipage}
\begin{minipage}[b]{0.5\linewidth}
\centering
\includegraphics[width=8cm]{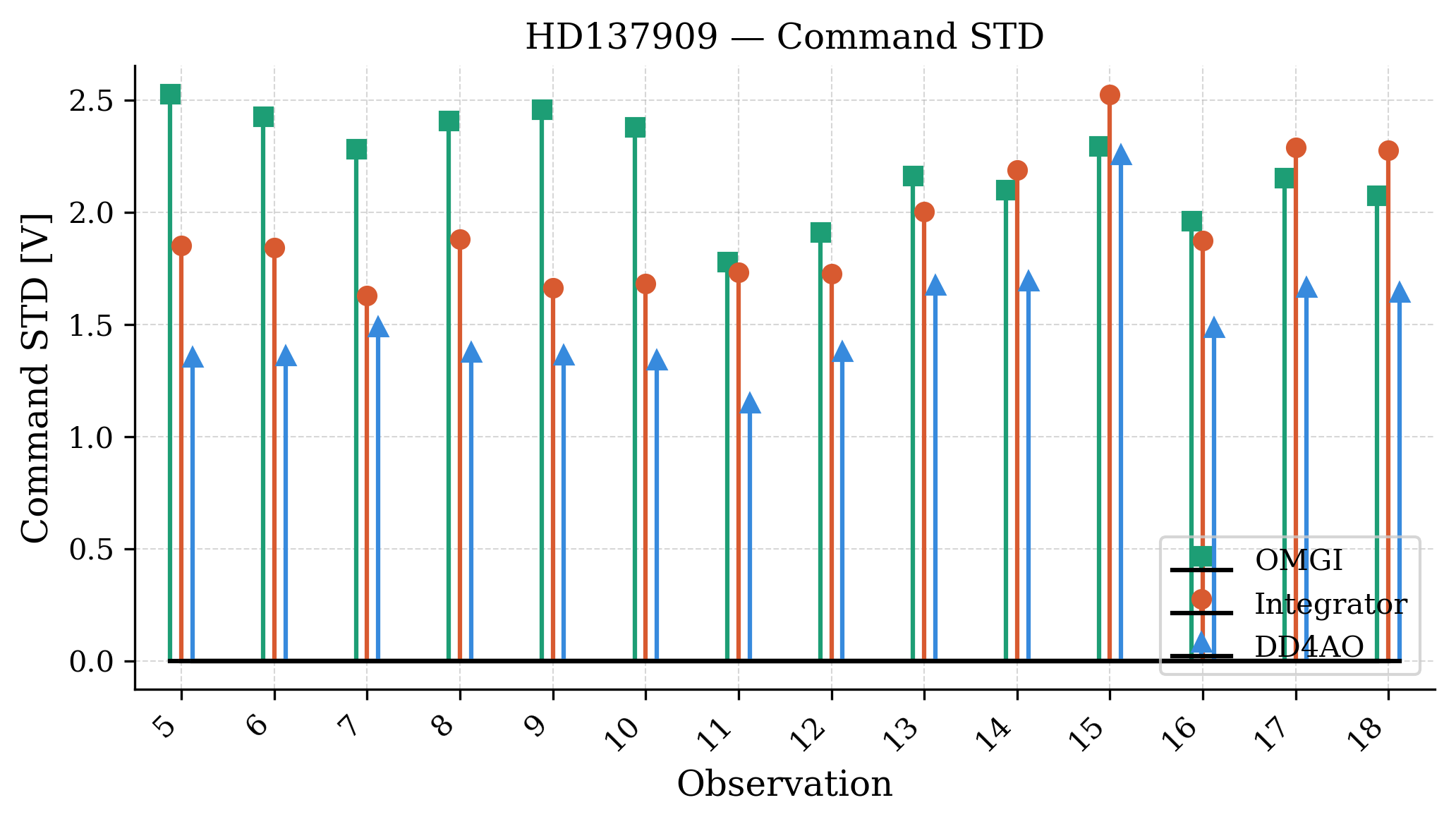}
\captionsetup{width=8cm}
\caption{Command standard deviation across 14 HD137909 observations, comparing 
DD4AO, the integrator, and OMGI.}
\label{fig:hd_cmd}
\end{minipage}
\end{figure}

Figure~\ref{fig:hd_res} shows the residual RMS from the reconstructed KL modes 
across 14 observations taken within a one-hour period. DD4AO consistently outperforms the integrator, though the 
margin is not statistically significant, which is attributable to the lower 
wavefront sensor SNR on this fainter star.
Figure~\ref{fig:hd_cmd} shows the command standard deviation in volts for the same 
observations. In this case, DD4AO uses notably less stroke than the other 
controllers, while OMGI appears more aggressive.

\begin{figure}[H]
\begin{minipage}[b]{0.5\linewidth}
\centering
\includegraphics[width=8cm]{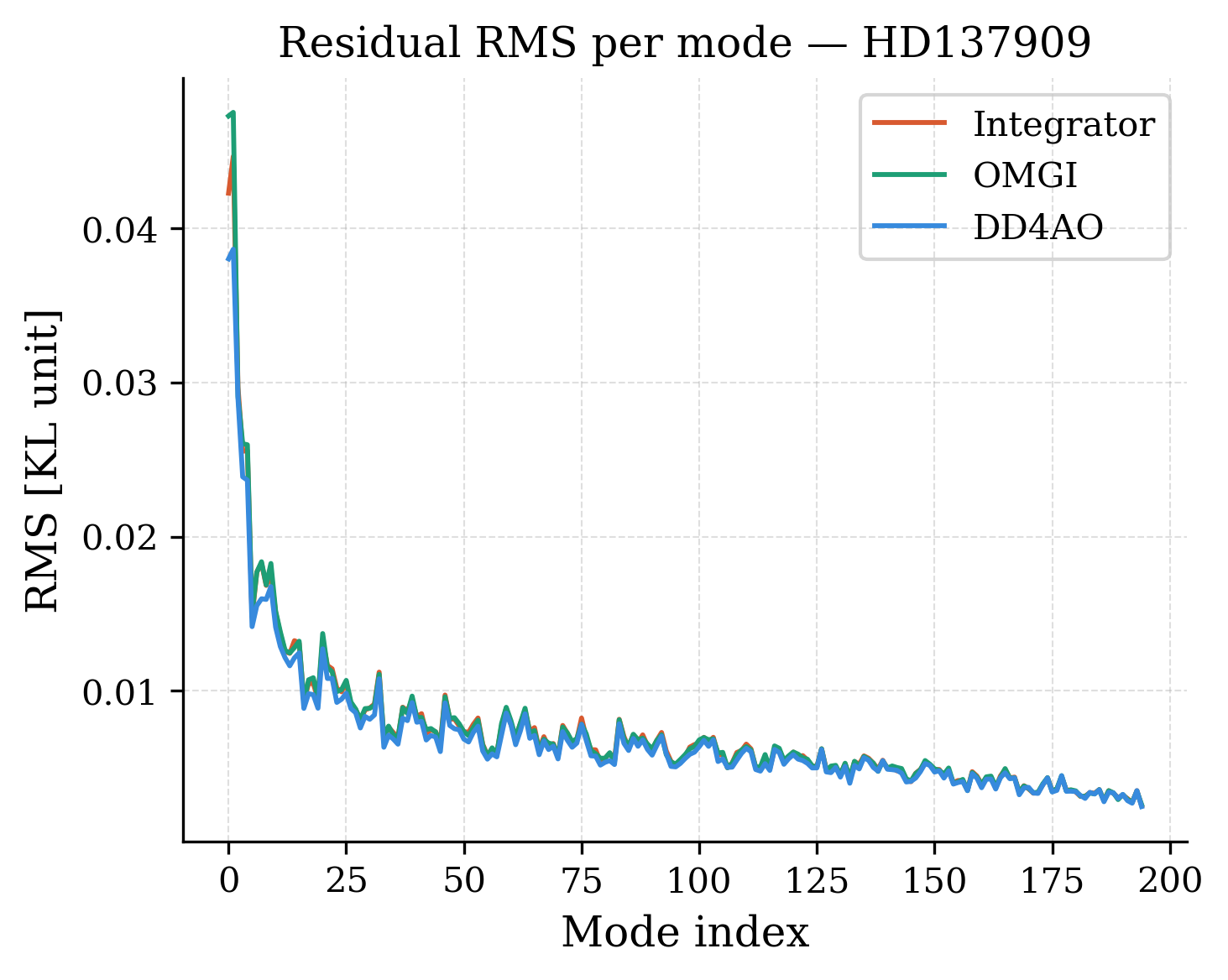}
\captionsetup{width=8cm}
\caption{Residual RMS per KL mode for a representative HD137909 exposure, 
comparing DD4AO, the integrator, and OMGI.}
\label{fig:hd_rms_mode}
\end{minipage}
\begin{minipage}[b]{0.5\linewidth}
\centering
\includegraphics[width=8cm]{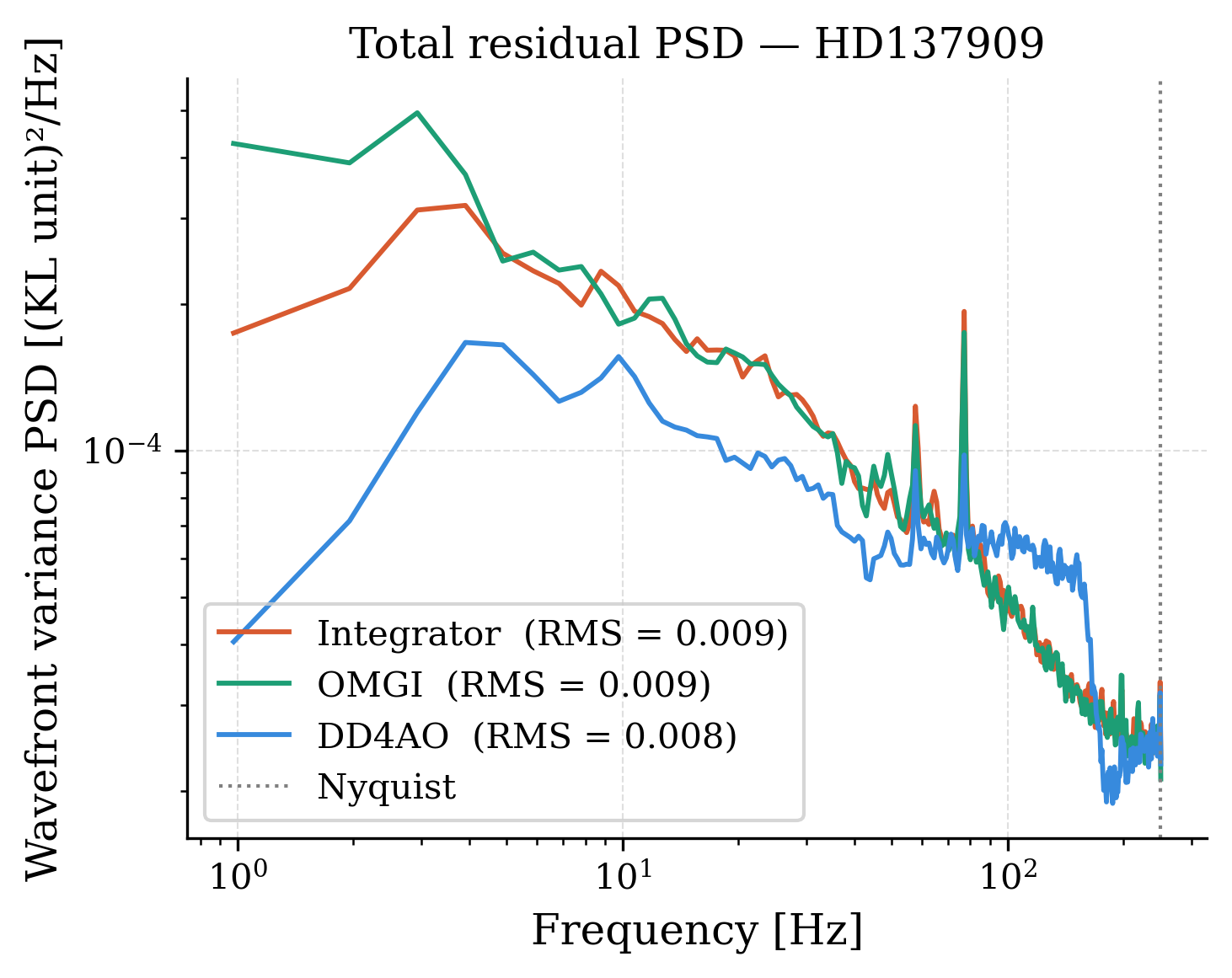}
\captionsetup{width=8cm}
\caption{Total residual PSD for a representative HD137909 exposure, comparing 
DD4AO, the integrator, and OMGI.}
\label{fig:hd_psd}
\end{minipage}
\end{figure}

Figures~\ref{fig:hd_rms_mode} and~\ref{fig:hd_psd} show, for one representative 
observation, the residual RMS per mode and the total residual PSD, respectively. 
As expected given the low SNR, there is little difference in performance between 
the control laws, except for the first two modes. The residual PSD further reflects 
the low signal level of this faint star. Notably, the two vibration peaks visible 
in the integrator and OMGI PSDs are damped when using DD4AO.

\begin{figure}[H]
\begin{minipage}[b]{0.33\linewidth}
\centering
\includegraphics[width=5.5cm]{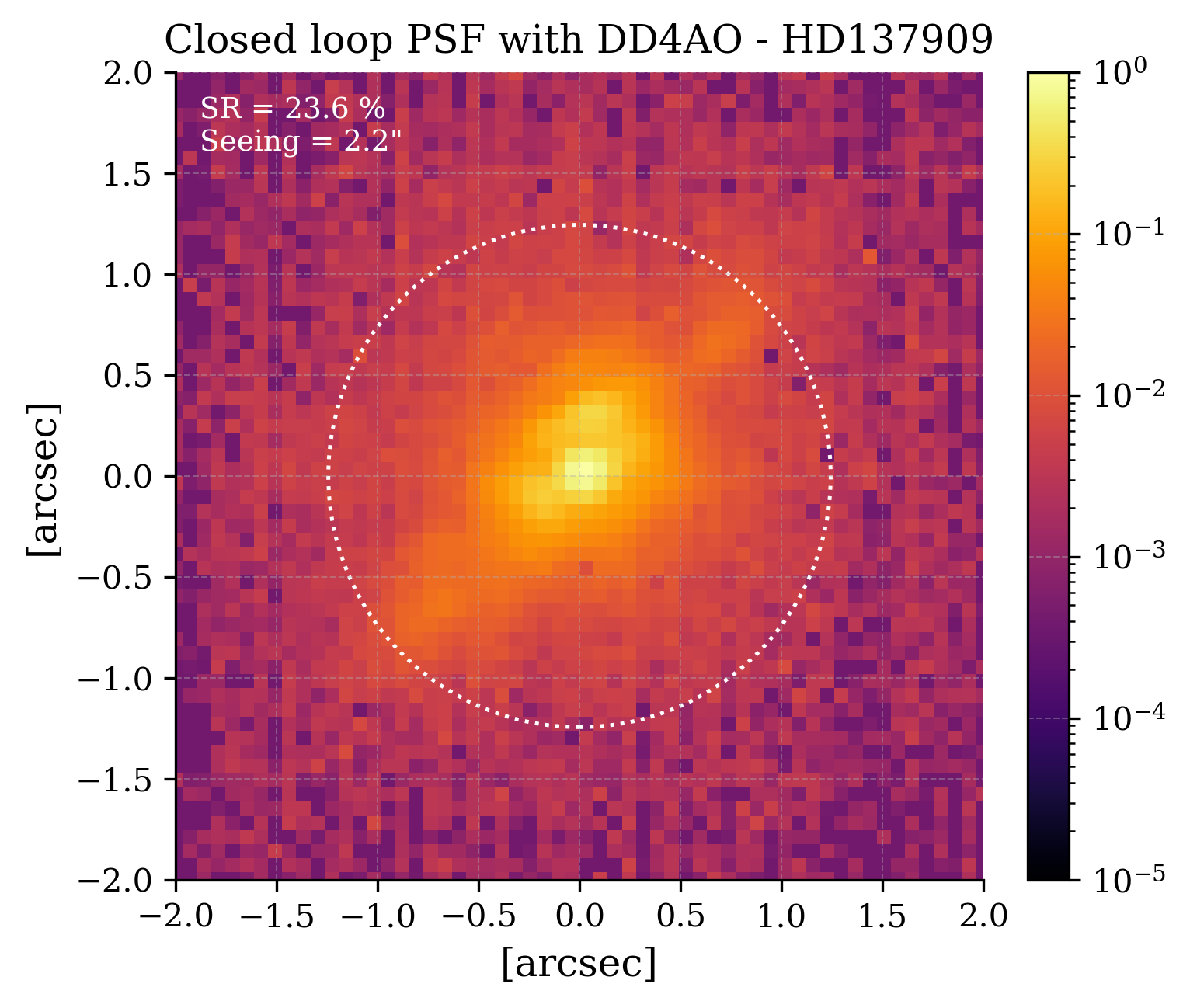}
\captionsetup{width=5.5cm}
\caption{Closed-loop PSF on HD137909 using DD4AO. Strehl ratio at 
$\lambda = 1310\,\text{nm}$: 23.6\%.}
\label{fig:psf_dd4ao_hd}
\end{minipage}
\begin{minipage}[b]{0.33\linewidth}
\centering
\includegraphics[width=5.5cm]{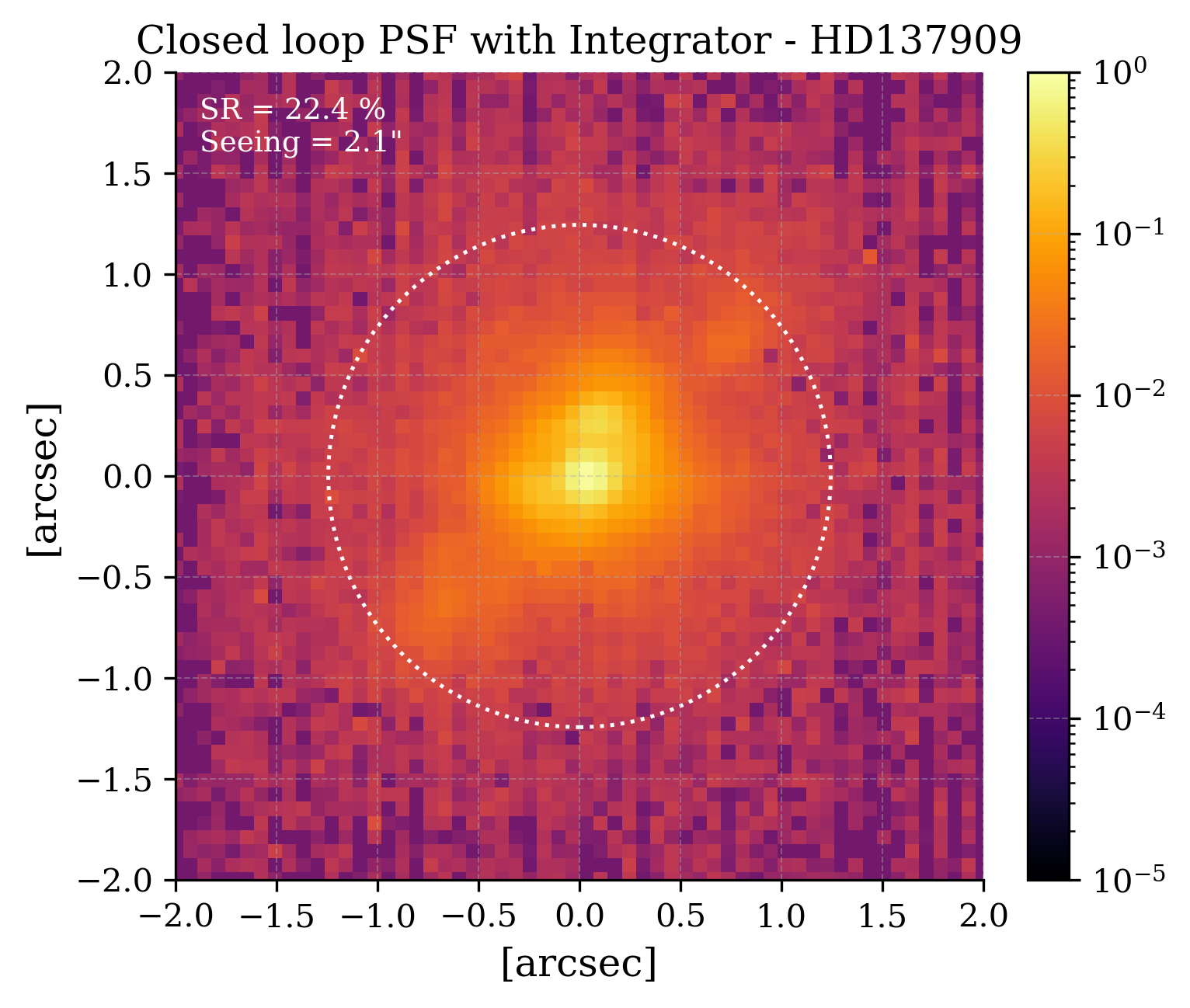}
\captionsetup{width=5.5cm}
\caption{Closed-loop PSF on HD137909 using the integrator. Strehl ratio at 
$\lambda = 1310\,\text{nm}$: 22.4\%.}
\label{fig:psf_int_hd}
\end{minipage}
\begin{minipage}[b]{0.33\linewidth}
\centering
\includegraphics[width=5.5cm]{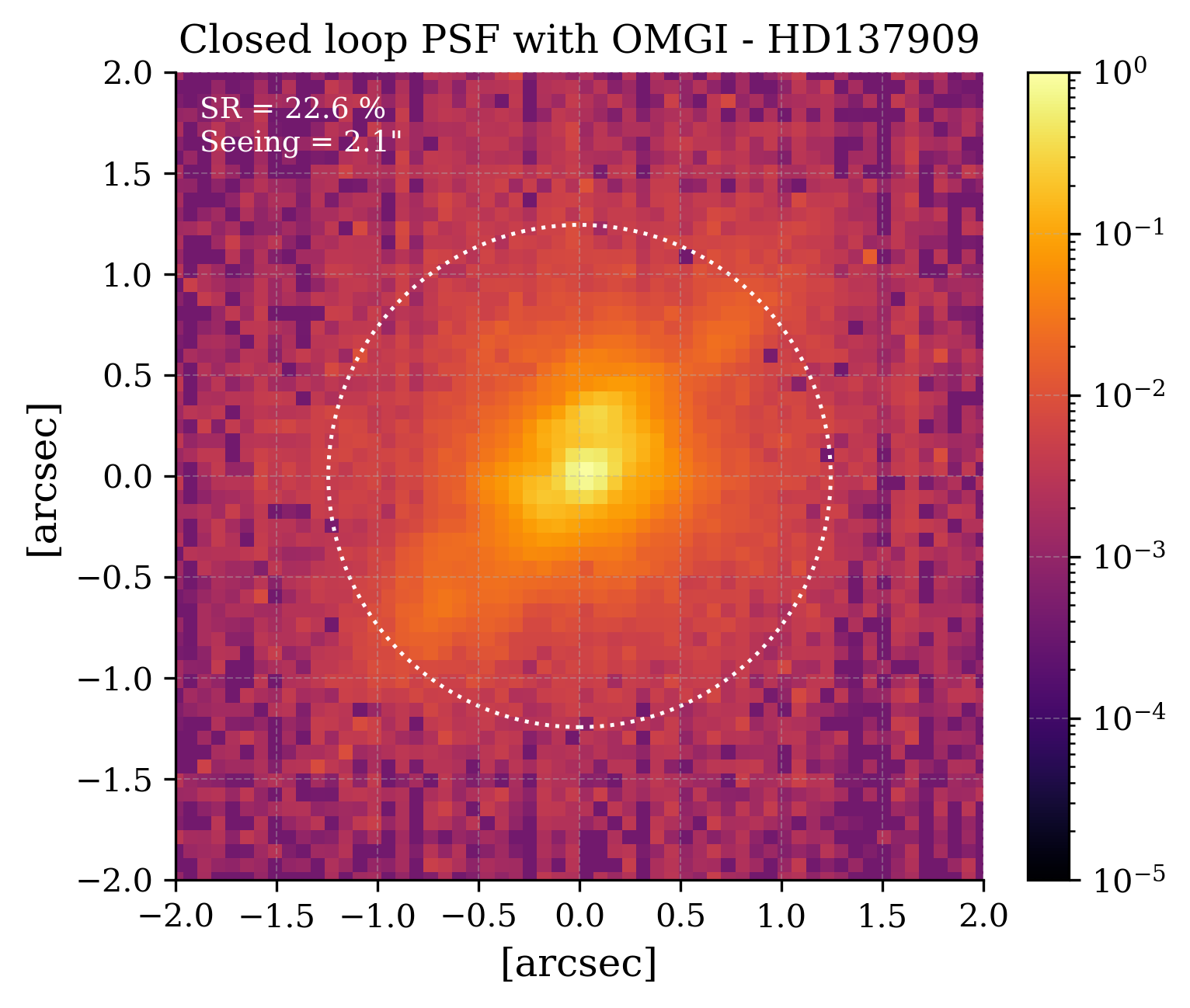}
\captionsetup{width=5.5cm}
\caption{Closed-loop PSF on HD137909 using OMGI. Strehl ratio at 
$\lambda = 1310\,\text{nm}$: 22.6\%.}
\label{fig:psf_omgi_hd}
\end{minipage}
\end{figure}

Finally, Figures~\ref{fig:psf_dd4ao_hd},~\ref{fig:psf_int_hd} 
and~\ref{fig:psf_omgi_hd} show the closed-loop PSFs at 
$\lambda = 1310\,\text{nm}$ for a representative observation. Consistent with the 
residual RMS results, the Strehl ratios are comparable across all three control 
laws, with differences falling within the measurement uncertainty expected for this 
SNR regime. The strehl ratio is estimated with maoppy \cite{maoppy}.

\section{CONCLUSION}

This paper presented the on-sky validation of DD4AO, a novel frequency-domain, 
data-driven controller for adaptive optics, conducted using the PAPYRUS instrument 
at the Observatoire de Haute-Provence. The results demonstrate that DD4AO 
successfully maintained a closed and stable loop over hour-long observations while 
continuously adapting to evolving atmospheric conditions, with instantaneous 
switching between controllers enabling direct statistical comparisons throughout 
each exposure.

On the bright star Arcturus, DD4AO consistently outperformed the classical 
integrator in terms of residual wavefront RMS, with the improvement concentrated 
in the low-order modes and low temporal frequencies. This gain was accompanied by 
a 5\% increase in Strehl ratio at $\lambda = 1310\,\text{nm}$, from 28.8\% to 
33.9\%. On the fainter star HD137909, performance differences between DD4AO, the 
integrator, and OMGI were smaller, as expected in the low-SNR regime. Nevertheless, 
DD4AO demonstrated notably lower actuator stroke usage across all 14 observations 
and effective suppression of vibration peaks that remained visible in the residual 
PSDs of both the integrator and OMGI.

Across both targets, DD4AO achieved equal or better wavefront correction than the 
standard controllers while requiring less deformable mirror stroke, a result with 
direct implications for actuator longevity and sky coverage in future instruments. 
These results validate DD4AO as a robust and adaptive control solution under real 
observing conditions, and represent an important milestone towards its 
implementation in the next-generation XAO instruments RISTRETTO and SAXO+ at 
the VLT.

Future work will focus on improving the scalability of the real-time pipeline for 
deployment in XAO instruments controlling a large number of modes. In particular, 
the hard real-time control process will be further optimized to maintain 
integrator-like latency as the number of controlled modes scales up. 
On-sky validation will also be extended to a broader range of atmospheric conditions 
and guide star magnitudes.

\bibliography{report} 
\bibliographystyle{spiebib} 

\end{document}